\documentclass[a4paper,fleqn,usenatbib]{mnras}

\usepackage[T1]{fontenc}
\usepackage{ae,aecompl}

\usepackage{graphicx}	
\usepackage{amsmath}	
\usepackage{amssymb}	

\def\msun{{\rm\,M_\odot}}

\def\gtsima{$\; \buildrel > \over \sim \;$}
\def\simgt{\lower.5ex\hbox{\gtsima}}

\def\msun{\hbox{M$_\odot$}}
\title[A stripped star cluster from the SMC]{First evidence of a stripped 
star cluster from the Small Magellanic Cloud}

\author[A.E. Piatti \& S. Lucchini]{
Andr\'es E. Piatti$^{1,2}$\thanks{E-mail: andres.piatti@unc.edu.ar} and
Scott Lucchini$^{3}$\\
$^{1}$Instituto Interdisciplinario de Ciencias B\'asicas (ICB), CONICET-UNCUYO, 
Padre J. Contreras 1300, M5502JMA, Mendoza, Argentina\\
$^{2}$Consejo Nacional de Investigaciones Cient\'{\i}ficas y T\'ecnicas, Godoy Cruz 
2290, C1425FQB,  Buenos Aires, Argentina\\
$^{3}$Department of Physics, University of Wisconsin–Madison, Madison, WI, USA
}

\date{Accepted XXX. Received YYY; in original form ZZZ}

\pubyear{2022}

\begin{document}
\label{firstpage}
\pagerange{\pageref{firstpage}--\pageref{lastpage}}
\maketitle

\begin{abstract}
We present results on the recently discovered stellar system YMCA-1,
for which physical nature and belonging to any of the Magellanic System
galaxies have been irresolutely analyzed. We used SMASH and {\it Gaia} 
EDR3 data sets to conclude that we are dealing with a small star cluster. Its reddening free,
field star decontaminated colour-magnitude diagram was explored in order
to obtain the cluster parameters. We found that YMCA-1 is a
small (435 $\msun$), moderately old (age = 9.6 Gyr), moderately metal-poor 
([Fe/H] = -1.16 dex) star cluster, located at a nearly Small Magellanic
Cloud (SMC) distance (60.9 kpc)  from the Sun, at $\sim$ 17.1 kpc to the East from the 
Large Magellanic Cloud (LMC) centre. The derived 
cluster brightness and size would seem to suggest some resemblance to the
recently discovered faint star clusters in the Milky Way (MW) outer halo, although
it does not match their age-metallicty relationship, nor those of MW
globular clusters formed in-situ  or ex-situ,
nor that of  LMC clusters either, but is in agreement with
that of SMC old star clusters. We performed numerical Monte Carlo 
simulations integrating its orbital motion backward in the MW-LMC-SMC system with 
radially extended dark matter haloes that experience dynamical friction, and by exploring different 
radial velocity (RV) regimes for YMCA-1. For RVs $\gtrsim$ 300 km/s,
the cluster remains bound to the LMC during the last 500 Myrs.
The detailed tracked kinematic of YMCA-1 suggests that its could have been
stripped by the LMC from the SMC during any of the close interactions between
both galaxies, a scenario previously predicted by numerical simulations.
\end{abstract} 




\section{Introduction}

The search for old star clusters in the outskirts of the Large Magellanic Cloud
(LMC) has long been motivated by the fact that they have been thought to be
key to reconstructing the early galaxy formation and chemical enrichment history
\citep{getal97}. As far as we are aware, a handful of faint star clusters (DES~4, 
DES~5, Torrealba~I, {\it Gaia~3}) have been recently discovered 
\citep[][and references therein]{torrealbaetal2019,bicaetal2020}, in addition to other three
stellar systems of uncertain nature, namely,
SMASH~1 \citep{martinetal2016}, DELVE~2 \citep{cernyetal2021}, and YMCA-1 
\citep{gattoetal2021}. They are stellar aggregates that
can be either ultra-faint dwarf galaxies or old star clusters. 
There are others $\sim$ 22 objects discovered to date in the outskirts of the
Magellanic Cloud galaxies that resulted to be ultra-faint dwarf galaxies.

According to \citep{martinetal2016}, SMASH-1 could be an ancient, very elongated, 
star cluster that is being tidally disrupted by the LMC at a distance of 
$\sim$13 kpc from it. They suggested that the stellar system could have been stripped
from the SMC, based on the simulations by \citep{carpinteroetal2013}.
Nevertheless, among 
the known LMC globular clusters, NGC~1841 tightly matches 
the age and metallicity of SMASH~1. 
The former is located at a deprojected distance of 14.25 kpc 
from the LMC centre, toward the same southern LMC region where SMASH~1 is located, 
and shows an LMC disk kinematics \citep{piattietal2019}. This suggests that SMASH~1 could be
a member of the LMC globular cluster population.
DELVE~2 is located at 28 kpc from the LMC toward the Small Magellanic Cloud (SMC)
and, as opposed to SMASH~1, is a nearly spherical very old stellar system. However, simulations of 
the accreted satellite population of the LMC led \citet{cernyetal2021} to conclude that it 
is very likely associated with the LMC.

YMCA-1 has been suggested to be a remote Milky Way (MW) halo star cluster located at 100 kpc from
the Galactic centre. In the sky, it is projected $\sim$ 13$\degr$ to the East of the LMC centre. 
However, \citet{gattoetal2021} pointed out that the stellar system nature and the object distance
need to be confirmed from deeper photometry. Precisely, the main aim of this work is to
take advantage of the SMASH database (obtained similarly as that by the DELVE's team; 
\citet{drlicawagneretal2021}) to revisit YMCA-1, in order to estimate accurate astrophysical
properties and to uncover its origin. In Section 2 we make use of the available SMASH data sets
to estimate YMCA-1's distance, age, metallicity and present mass, and to derive its
projected half-light radius from its surface density profile. Section 3 discusses the
resulting fundamental parameters to conclude on the stellar system nature and its origin.

\begin{figure*}
\includegraphics[width=\textwidth]{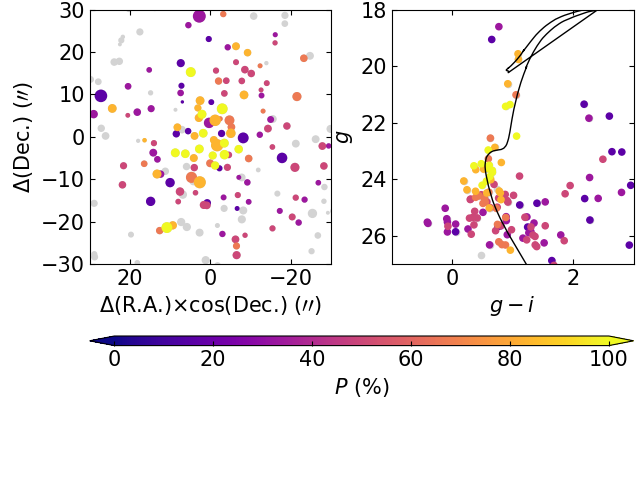}
\caption{ {\it Left panel:}  Chart of all the measured stars in the field of YMCA-1. 
The size of the symbols is proportional to the $g$ brightness of the star. {\it Right panel:}
Colour-magnitude diagram of YMCA-1. The isochrone which best represents the
distribution of stars with $P >$ 50 per cent is overplotted. Filled circles in both
panels are colour-coded according to the assigned membership probability $P$.}
\label{fig1}
\end{figure*}

\section{Data handling}

Among the public surveys that cover the Magellanic System,
the Survey of the MAgellanic Stellar History \citep[SMASH,][]{nideveretal2021}
is a suitable one for our purposes. Its limiting magnitude is beyond the 25th
magnitude in the outskirts of the Magellanic Clouds, which is $\sim$1.5 mag
underneath the YMCA-1 turnoff \citep[see Fig. 1 in][]{gattoetal2021}.
We retrieved data from the Astro Data 
Lab\footnote{https://datalab.noirlab.edu/smash/smash.php}, which is part of the 
Community Science and Data Center at NSF’s
National Optical Infrared Astronomy Research Laboratory. They consist of R.A and 
Dec. coordinates, PSF $g,i$ magnitudes and their respective errors, $E(B-V)$ 
interstellar reddening and $\chi$ and {\sc sharpness} parameters of stellar sources
located inside a radius of 5$\arcmin$ from the YMCA-1's centre. According to
\citet{gattoetal2021}, the object has a radius of 0.3$\arcmin$. We selected those
sources with 0.2 $\le$ {\sc sharpness} $\le$ 1.0 and $\chi^2$ $<$ 0.5, in order to exclude 
bad pixels, cosmic rays, galaxies, and unrecognized double stars. The photometric 
performance is as mentioned in \citet{nideveretal2017a}. In their Table 4, they
presented the SMASH average photometric transformation equations. We added in
quadrature zero-point, extinction and colour term errors, and computed an 
accuracy $\la$ 0.02 mag in $gi$. They showed that these calibration errors imply a
SMASH photometry precision of $\sim$ 0.5-0.7 per cent in $gi$. Such a precision
implies in turn an uncertainty of $\sim$ 0.12-0.17 mag in $gi$, for a star at
$g \ga$ 25.0 mag. 

Aiming at disentangling the true nature of YMCA-1, its observed colour-magnitude diagram
(CMD) needs to be cleaned from field star contamination. The contamination of field 
stars plays an important role, because it is not straightforward  to consider a star as 
a member of the object of interest only on the basis of  its position in that CMD. 
We cleaned the YMCA-1's CMD using the photometry of a reference star field 
placed far from the object field, but not too far from it as to become unsuitable 
as representative of the star field projected along the line-of-sight (LOS) of 
YMCA-1. Even though YMCA-1 is located $\sim$13$\degr$ to the East of the LMC centre,
and therefore is not projected onto a crowded star field or is not affected by differential
reddening, it is highly possible to find differences between the stellar density, magnitude 
and colour distributions of different adjacent star fields surrounding YMCA-1.
For this reason we decided to clean a circular area of radius 0.5$\arcmin$ centred on 
YMCA-1, using at a time six different circular areas with the same radius distributed
around the YMCA-1's circle. We thus increase the statistics of the cleaning procedure.

The method employed to remove field stars was devised by  \citet{pb12}, which was 
satisfactorily applied elsewhere 
\citep[e.g.,][and references therein]{petal2018,piatti2021d}. The procedure
proved to be  successful for clusters projected on to crowded fields and affected by 
differential reddening \citep[see, e.g.,][and references therein]{pft2020}.
It uses the magnitude and colour of each star in the reference
star field CMD and finds the closest one in magnitude and colour in the YMCA's CMD and
subtracts it. The methodology to select stars to subtract from the YMCA-1's CMD
consists in defining boxes centred on the magnitude and colour of each star of the
reference star field; then to superimpose them on to the YMCA's CMD, and finally to choose 
one star per box to subtract. We started with
boxes with size of ($\Delta$$g_0$,$\Delta$$(g-i)_0$) = (1.00 mag, 0.25 mag)
centred on the ($g_0$, $(g-i)_0$) values of each reference field star,
in order to guarantee to find a star in the YMCA'a CMD with
the magnitude and colour within the box boundary. The reddening corrected $g_0$,$i_0$
magnitudes were derived by using the observed $g,i$ magnitudes, $E(B-V)$ values provided 
by SMASH and the $A_\lambda$/$E(B-V)$ ratios, for $\lambda$ =
$g,i$, given by  \citet{abottetal2018}. In the case that more than one star 
is located inside a box, the closest one to the centre of 
that (magnitude, colour) box is subtracted. The magnitude and colour errors of the stars 
in the YMCA-1's CMD were taken into account while searching for a 
star to be subtracted. With that purpose, we allowed the search of stars in the YMCA-1's
CMD to vary within an interval of 
$\pm$1$\sigma$, where $\sigma$ represents the errors in their magnitude and colour, 
respectively. We allowed up to 1000 random combination of their magnitude and colour
errors.

The outcome of the cleaning procedure
is a YMCA-1's CMD that likely contains only members. We finally assigned a membership 
probability to each star that remained unsubtracted
after the decontamination of the YMCA-1's CMD. Because we produced six different cleaned
CMDs (one per reference star field employed), we defined the probability
$P$ ($\%$) = 100$\times$$N$/6, where $N$ represents the number of times a
star was not subtracted during the six different CMD cleaning executions. With that
information on hand, we built Fig.~\ref{fig1}, which shows the spatial distribution
and the CMD of all the measured stars located in the field of YMCA-1. Stars with
different $P$ values were plotted with different colours.
Fig.~\ref{fig1} reveals a red giant branch and a Main Sequence turnoff that are 
clearly highlighted as the most probable YMCA-1 sequences. From a total of 125 stars
located in a circle of radius 0.5$\arcmin$, 46 resulted with $P$ $>$ 50 per cent.

We fitted theoretical isochrones  computed by 
\citet[][PARSEC\footnote{http://stev.oapd.inaf.it/cgi-bin/cmd}]{betal12} for the
SMASH photometric system,
We used PARSEC v1.2S isochrones spanning metallicities (log($Z/Z_\odot$)) from 0.00005 
dex up to 0.001 dex, in steps of 0.001 dex and log(age /yr) from 9.5 up to 10.1 in steps 
of 0.025. We fitted these isochrone sets to stars with $P >$ 50 per cent allowing the
true distance modulus to vary between 18.0 mag (40 kpc) and 20.0 mag (100 kpc). We 
employed routines of the 
Automated Stellar Cluster Analysis code \citep[\texttt{ASteCA,}][]{pvp15}
that allowed us to derive simultaneously the metallicity, the age and the distance of 
YMCA-1. \texttt{ASteCA} is a suit of tools that produces a synthetic CMD that best 
matches the cleaned star cluster CMD. The metallicity, age, distance, reddening, star 
cluster present mass and binary fraction associated to that generated synthetic CMD were 
adopted as the best-fitted star cluster properties. Because individual 
$E(B-V)$ values are available from the SMASH data sets, we entered into \texttt{ASteCA}
with the intrinsic YMCA-1's CMD.

For generating the synthetic CMDs, we adopted the initial mass function of \citet{kroupa02}; 
a minimum mass ratio for the generation of binaries of 0.5. Star cluster mass and binary 
fractions were set in the ranges 100-5000 $\msun$ and 0.0-0.5, respectively. 
We explored the parameter space of the 
synthetic CMDs through the minimization of the likelihood function defined by 
\citet[][the Poisson likelihood ratio (eq. 10)]{tremmeletal2013} using a parallel tempering 
Bayesian MCMC algorithm, and the optimal binning \citet{knuth2018}'s method.
Errors in the obtained parameters are
estimated from the standard bootstrap method described in \citet{efron1982}. We refer 
the reader to the work of \citet{pvp15} for details concerning the implementation of 
these algorithms. The resulting fundamental parameters for YMCA-1 turned out to be:
distance = 60.9$^{\rm +14}_{\rm -12}$ kpc (true distance modulus = 
18.92$^{\rm +0.45}_{\rm -0.47}$ mag); age = 9.6$^{\rm +3.5}_{\rm -2.6}$ Gyr, 
[Fe/H] = -1.16$^{\rm +0.11}_{\rm -0.14}$ dex; and present mass = 435$\pm$157 $\msun$.
The isochrone for the resulting mean values is superimosed in Fig.~\ref{fig1}.

\begin{figure}
\includegraphics[width=\columnwidth]{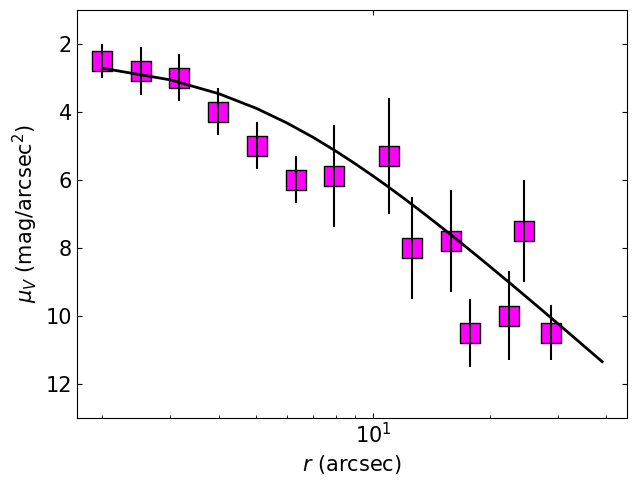}
\caption{Surface density profile built from stars with membership probability
$P >$ 50 per cent. The best-fittet \citet{plummer11}'s model 
($r_h$ = 6.5 arcsec, equivalent to 1.92 pc) is
overplotted.}
\label{fig2}
\end{figure}

We also used stars with $P >$ 50 per cent to build the surface density profile of YMCA-1.
In order to do that, we computed the integrated $M_V$ magnitude in annuli
of 0.02$\arcmin$, 0.04$\arcmin$, 0.06$\arcmin$, 0.08$\arcmin$ and 0.10$\arcmin$ 
wide, and then calculated their average and dispersion. The integrated $M_V$ magnitudes
were computed from the sum of individual stellar luminosities ($L_*$), that were in turn 
obtained by interpolating the reddening free $g_0$ magnitudes in the theoretical 
isochrone corresponding to the mean age and metallicity of YMCA-1, and the following
expression:

\begin{equation}
M_V = 4.72 - 2.5\times log(\sum L_*) + 0.18
\end{equation}

\noindent where 4.72 mag is the adopted bolometric magnitude of the Sun, and
-0.18 mag is the mean bolometric correction from \citet[][their figure 14]{masanaetal2006}
for a mean effective temperature of log($T_{eff}$)= 4.77$\pm$0.03, which
corresponds to the average of individual effective temperatures interpolated in the 
above  isochrone for stars with $P >$ 50 per cent. The resulting surface density profile 
with its respective uncertainties is depicted in Fig.~\ref{fig2}, 
where we superimposed the best-fitted \citet{plummer11}'s model (derived half-light radius
$r_h$ = 6.5$^{\rm +2.4}_{\rm -2.5}$ arcsec, equivalent to 
1.92$^{\rm +0.72}_{\rm -0.74}$ pc).  As can be seen, the surface density
profile is not smooth as those of  populous globular clusters. This is because
the number of stars is relatively small, and they are irregularly spatially distributed as
a function of their brightnesses, so that stochastic effects have a role. Such
stochastic effects are reflected in the $r_h$ uncertainty. We also took advantage of the above equation to
calculate the integrated $M_V$ mag of YMCA-1, which turned out to be -1.70$\pm$0.50 mag, 
where the uncertainty comes from using a Monte Carlo sampling of the involved parameters and 
their uncertainties.

\begin{figure}
\includegraphics[width=\columnwidth]{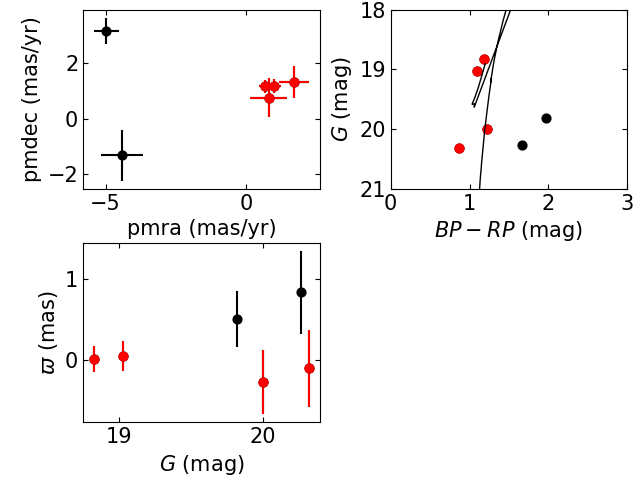}
\caption{Relationships between {\it Gaia} EDR3 retrieved parameters for stars
located within a circle of radius 0.5$\arcmin$ from the YMCA-1's centre. We 
represented highly probable members with filled circles. The theoretical isochrone for the
adopted mean distance, age, and metallicity is overplotted in the {\it Gaia}
CMD.}
\label{fig3}
\end{figure}

Finally, from the {\it Gaia} Early Data Release\footnote{https://archives.esac.esa.int/gaia.}
database \citep[{\it Gaia} EDR3;][]{gaiaetal2016,gaiaetal2020b}, we extracted
parallaxes ($\varpi$), proper motions in right ascension (pmra) and declination (pmdec),
excess noise (\texttt{epsi}), the significance of excess of noise (\texttt{sepsi}), and $G$,
$BP$, and $RP$ magnitudes for stars located within 0.5$\arcmin$ from the YMCA-1's centre.
We pruned the data with \texttt{sepsi} $<$ 2 and \texttt{epsi} $<$ 1, which define a good balance 
between data quality and the number of retained objects for our sample \citep[see also][and references therein]{piatti2021b}.
We then built the vector point diagram, the CMD, and the parallax versus magnitude diagram
(see Fig.~\ref{fig3}) for all the retrieved stars. As can be seen, four stars drawn with
red filled circles seem to be highly probable YMCA-1's members, because of their similar
proper motions, of their placement along the red giant branch, and of their parallaxes. For 
them, we  obtained  $<$pmra$>$ = 1.044$\pm$0.402 mas/yr and $<$pmdec$>$= 1.107$\pm$0.209 
mas/yr. Table~\ref{tab:tab1} summarizes the derived YMCA-1's properties.

\begin{table}
\caption{Derived properties of YMCA-1}
\label{tab:tab1}
\begin{tabular}{@{}lr}\hline
Property & value \\\hline
$\alpha$$_{\rm 2000}$ & 110.8378$\degr$ \\
$\delta$$_{\rm 2000}$ &-64.8319$\degr$ \\
$r_h$ & 1.92$^{\rm +0.72}_{\rm -0.74}$ pc \\
$(m-M)_0$ & 18.92$^{\rm +0.45}_{\rm -0.47}$ mag \\
distance$_{MW}$ & 60.9$^{\rm +14}_{\rm -12}$ kpc \\
distance$_{LMC}$ & $\sim$ 17.1 kpc \\ 
distance$_{SMC}$ & $\sim$ 86.0 kpc \\
Age & 9.6$^{\rm +3.5}_{\rm -2.6}$ Gyr \\
{\rm [Fe/H]} & -1.16$^{\rm +0.11}_{\rm -0.14}$ dex \\
Mass & 435$\pm$157 $\msun$ \\
$<$log($T_{eff}$)$>$ & 4.77$\pm$0.03 \\
$M_V$ & -1.70$\pm$0.50 mag \\
$<$pmra$>$ & 1.044$\pm$0.402 mas/yr \\
$<$pmdec$>$ & 1.107$\pm$0.209 mas/yr \\\hline
\end{tabular}
\end{table}

\section{Analysis and discussion}

\begin{figure*}
\includegraphics[width=\columnwidth]{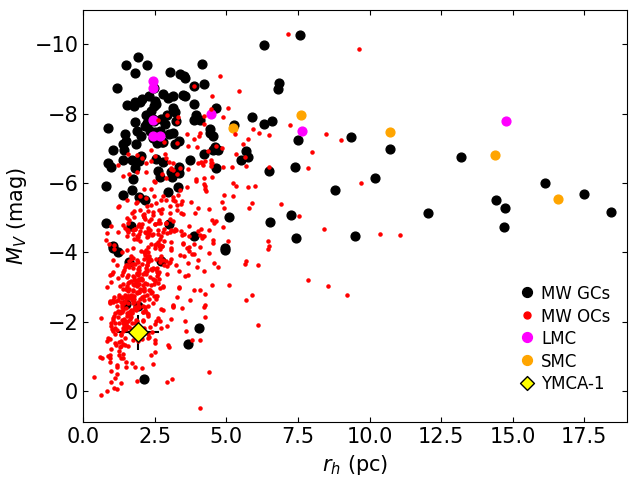}
\includegraphics[width=\columnwidth]{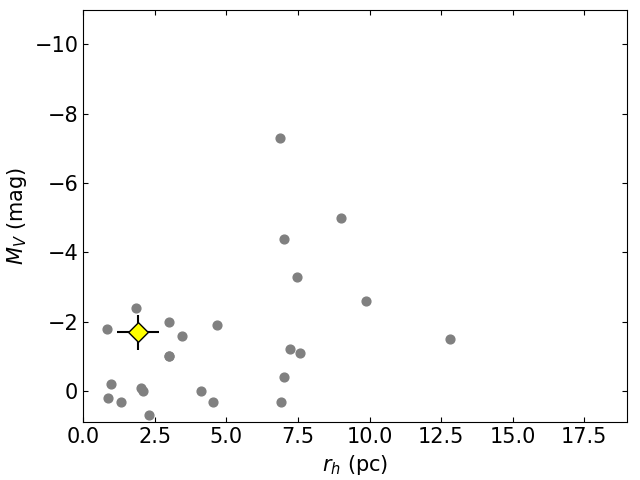}
\includegraphics[width=\columnwidth]{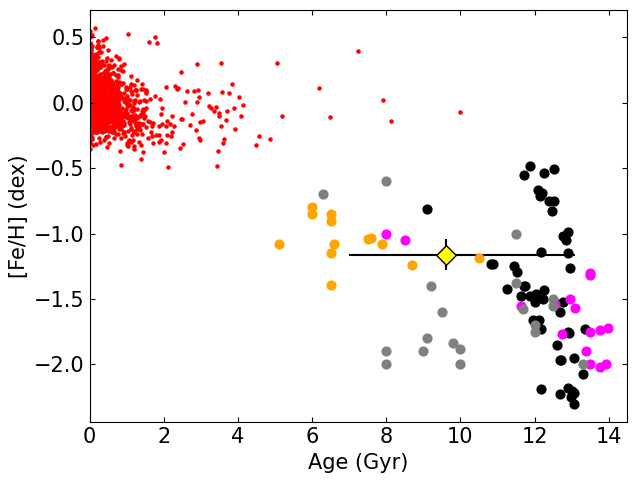}
\caption{{\it Left-top panel:} $M_V$ versus $r_h$ relationship for MW globular clusters 
\citep[black filled circles;][]{harris1996,kruijssenetal2019}; LMC and SMC 
globular clusters 
\citep[magenta and orange filled cicles, respectively;][]{getal09,mg2003a,baetal13,pm2018,piattietal2019,songetal2021,parisietal2022}; 
and MW
open clusters \citep[red points;][]{kharchenkoetal2009,kharchenkoetal2013,diasetal2021w}.
YMCA-1 is represented by a large filled yellow diamond, with its 
respective error bars. {\it Right-top panel}: same as left panel for recently discovered
faint star clusters in the MW halo: Balbinot~1 \citep{balbinotetal2013}, BLISS~1 
\citep{mauetal2019}, DELVE~1 \citep{mauetal2020}, DELVE~2 \citep{cernyetal2021}, DES~1
\citep{luqueetal2016}, DESJ0111-1341 \citep{luqueetal2017}, DES~3 \citep{luqueetal2018},
DES~4 and 5 \citep{torrealbaetal2019}, Gaia~1 y 2 \citep{koposovetal2017}, Gaia~ 3 to 7
\citep{torrealbaetal2019}, Kim~1 \citep{kj2015}, Kim~2 \citep{kimetal15}, Kim~3 
\citep{kimetal2016}, Koposov~1 and 2 \citep{koposovetal2007}, Laevens~3 \citep{laevensetal2015},
Mu\~noz~1 \citep{munozetal2012}, PSOJ174.0675-10.8774 \citep{laevensetal2014}, PS1~1
\citep{torrealbaetal2019}, Segue~3 \citep{fadelyetal2011}, and To~1 \citep{torrealbaetal2019}.
{\it Bottom panel:} age-metallicity relationship. Symbols and references as in the top panels.} 
\label{fig4}
\end{figure*}

The brightness and size of YMCA-1 suggest that we are dealing with a small star 
cluster. Indeed, Fig.~\ref{fig4},  top-left panel, shows that its integrated
absolute magnitude $M_V$ and projected half-light radius $r_h$ are placed in the
region of small MW open clusters, although it could also marginally belongs to
the family of recently discovered faint old star clusters of the outer halo of the
MW (see Fig.~\ref{fig4}, top-right panel).
\citet{gattoetal2022} obtained a 
larger half-light radius (3.5 pc, $\sim$ 2.2$\sigma$ larger) 
and fainter absolute magnitude (-0.47 mag, $\sim$2.5$\sigma$ fainter)
that make the object falls in  the middle among these faint star clusters
\citep{torrealbaetal2019}. 

Although there are very few MW open clusters with an age similar to that of YMCA-1
\citep{andersetal2021}, none of them is as metal deficient as [Fe/H] $\sim$ -1.2 dex;
the most metal-poor open clusters having [Fe/H] $\sim$ -0.5 dex \citep{diasetal2021w}.
This implies that YMCA-1 departs significantly from the age-metallicity relationship of 
the MW disk, and hence we conclude that it could not have formed in the Milky Way.
YMCA-1 would
not seem to be part of the population of MW globular clusters formed in-situ 
either, because of its distinct age and metallicity (see upper sequence of
black filled circles in Fig.~\ref{fig4}, bottom
panel \citep{kruijssenetal2019}). 
Globular clusters associated with dwarf galaxy 
accretion events (e.g., Sagittarius, Gaia-Enceladus, Sequioa, Koala, etc),
are distributed along the lower sequence of black filled circles of Fig.~\ref{fig4},
bottom panel. YMCA-1 falls slightly below it, while according to \citet{forbes2020}, 
the age-metallicity relationship of the High-Energy group of globular clusters would
seem to be the closest to the YMCA-1 placement.
These distant globular clusters spend much of their lifetime in the outer MW halo,
so that they have experienced less severe mass loss due to tidal effects than those
in the inner Galactic regions \citep{piatti2019,piattietal2019b}. Their present masses
are a couple of orders of magnitude larger than YMCA-1's derived mass, making it unlikely that 
YMCA-1 is a member of this outer globular cluster population, although
the age-metallicity relationship shows some overlap when the age error is considered.

The similitude of YMCA-1 with the recently discovered faint old star clusters is much 
difficult to disentangle. Indeed, some of them follow the age-metallicity relationship
sequence of accreted globular clusters (from the oldest ages until $\sim$ 6 Gyrs ago;
see Fig.~\ref{fig4}, bottom panel), 
while others are  metal-poor star clusters ([Fe/H] $\la$ -1.4 dex) and as old as YMCA-1.
None of these two sequences of old clusters would clearly seem to be that which YMCA-1 
belongs to.
As can be seen in Fig.~\ref{fig4}, bottom panel, YMCA-1 is in between them. Nonetheless,
because of the age uncertainty of YMCA-1, the above scenarios should not be ruled out
as possible origin of YMCA-1.

Having examined the possibility of YMCA-1 to be a MW star cluster, formed in-situ or
ex-situ, we analyse now whether it could have been born in the Magellanic Clouds. 
The LMC is known to harbour a population of $\sim$ 15 globular clusters older than
$\sim$ 12 Gyr and with metallicities more deficient than [Fe/H] $\sim$ -1.3 dex 
\citep{piattietal2019}. The absence of star clusters with ages between $\sim$ 4 and
11 Gyr, with the sole exception of ESO~121-SC03 and KMHK~1592 \citep{piatti2022},
have been extensively reported in the literature  (see Fig~\ref{fig4}, bottom panel). 
In the SMC there are some five
star clusters with ages between {\bf $\sim$ 7 and 12 Gyr }and metallicities in the
range  -1.2 $\la$ [Fe/H] (dex) $\la$ -1.0 \citep[][see also Fig.~\ref{fig4} here]{p11b}. 
Star clusters and field stars in the 
LMC and the SMC have similar age-metallicity relationships \citep{narlochetal2021,piatti2021f}, 
so that we would expect star clusters formed at any age accompanying the field star 
formation, which has not been interrupted in both galaxies since their formation
\citep{pg13,mazzietal2021}. This is not the case of LMC star cluster population, where 
we note the existence of only two star clusters of 8-9 Gyr old in the so-called LMC age gap
 \citep{piatti2022}. 
For this reason, we think that an SMC origin could explain the appearance of YMCA-1 in the 
outskirts of the LMC, at a distance of $\sim$ 17.1 kpc from its centre, unless an enough 
large population of 4-11 Gyr old clusters is discovered in the LMC.  

Recently, \citet{gattoetal2022} obtained deep VLT photometry with the aim of
disentangling the nature of YMCA-1 and determining its fundamental parameters. Because
of their resulting distance (55.5 kpc), they concluded that the object is
associated to the LMC, although a MW satellite is not discarded either.
According to them, YMCA-1 is placed in a transition region of the $M_V$ versus $r_h$ plane,
between ultra-faint dwarf galaxies and open clusters. They estimated an $r_h$ of 3.5 pc
from the fit of a smooth stellar number density profile  (see their Figure~1, right panel).
We note that a circle of radius 3.5 pc (12.9$\arcsec$)
nearly encompasses the entire cluster main body (see their Fig.~1, middle panel).
For these reasons we adopted our present $r_h$ value, which supports the existence of a
small star cluster. On the other hand, \citet{gattoetal2022} used as priors for determining
the cluster astrophysical parameters ages larger than 10 Gyr and distances smaller than
the value derived in \citet{gattoetal2021},  which constrained the fitting procedure.
We used wider ranges of ages and distances as \texttt{ASteCA}'s inputs and obtained
parameters in an overall agreement with those of \citet{gattoetal2022} and larger
uncertainties. We think that the latter are more realistic.

The tidal interaction between both Magellanic Clouds \citep[][among others]{wanetal2020,ruizlaraetal2020,mazzietal2021,williamsetal2021,romanduvaletal2021,
gradyetal2021,shippetal2021,cullinaneetal2022} can also be a source for star clusters
formed in the SMC were then stripped off by the LMC. Such a scenario was explored
numerically by \citet{carpinteroetal2013}, who modelled the dynamical interaction between 
both galaxies and their corresponding stellar cluster populations. They found that 
nearly 15 per cent of the SMC star clusters are captured by the LMC with eccentricities 
of the orbit of the SMC around the LMC $\ga$ 0.4, while another 20-50 per cent
is scattered into the intergalactic environment. Consequently, star clusters that 
originally belonged to the SMC could more likely be found in the outskirts of the LMC.  
YMCA-1 could form in the SMC (its age and metallicity agree well with the older enhanced 
formation event in the SMC \citep{p11b,p12a} and later captured by the LMC.

In order to confirm such a hypothesis, we have integrated the orbit of YMCA-1 backwards 
in time with respect to the MW-LMC-SMC system. The galaxies were included as point 
particles representing their gaseous and stellar disks with radially extended dark matter 
(DM) halos following Hernquist profiles \citep{hernquist90} including the effects of dynamical
friction. The MW was included as a live particle (not a static potential) thus allowing for the 
impact of its reflex motion on the other galaxies' orbits  \citep{gomezetal2015}. The properties of these galaxies are 
listed in Table~\ref{tab:tab2}.

\begin{figure*}
\includegraphics[width=\textwidth]{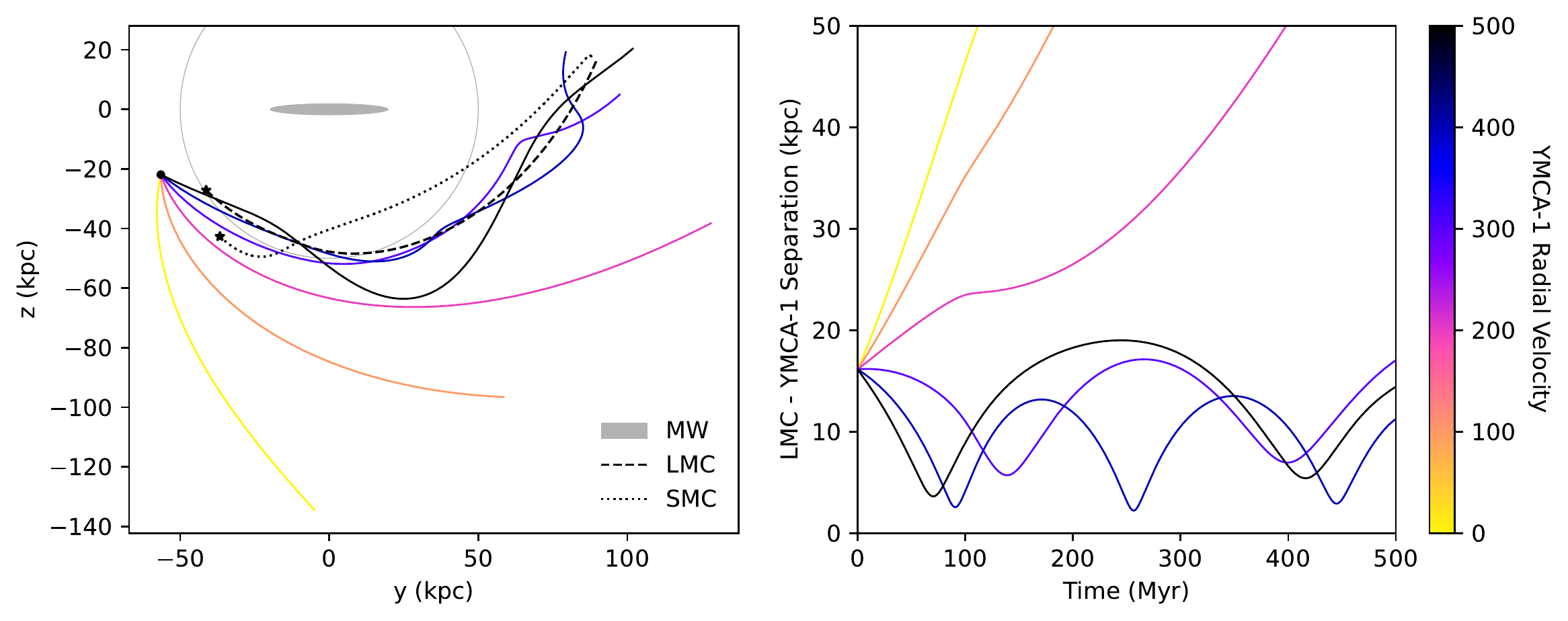}
\caption{Orbits of YMCA-1. The left panel shows the projection of the orbits of the LMC 
(dashed black line), SMC (dotted black line), and YMCA-1 (lines coloured by present-day 
RV) around the MW (grey oval) over the past 500 Myrs in the $y$-$z$ plane. In this 
coordinate system, the disk of the MW is assumed to be in the $x$-$y$ plane with the 
Sun's position at $(x,y,z)=(-8.1,0,0.02)$ kpc. The grey circle denotes 50 kpc away from 
the MW.}
\label{fig5}
\end{figure*}

We used the RA, Dec, distance, and proper motions of YMCA-1 to obtain 3D 
Galactocentric cylindrical coordinates using Astropy \citep{astropy2013,astropy2018} assuming 
a radial velocity (RV) ranging from 0 to 500 km/s. The orbits of the Magellanic Clouds around 
the MW with the various possible orbits of YMCA-1 are shown in Fig.~\ref{fig5}. Note that 
these orbits are being evolved backwards in time, so a positive RV (the cluster moving away 
from us at the present day) corresponds to the cluster moving closer to us in its recent past. 
For RVs greater than the LMC's observed value of 262.2 km s$^{-1}$ \citep{vdmareletal2002} 
the cluster is bound to the LMC and completes 2-3 orbits around it within the past 500 Myrs. 
This is the case even when integrating the orbits back 3 Gyr: YMCA-1 remains bound to the 
LMC if its RV $\gtrsim300$ km s$^{-1}$.

In addition to testing different RVs for the cluster, we also explored possible orbits within 
the uncertainties of the velocities of the LMC and SMC, as well as the uncertainties in the 
distance and proper motions of YMCA-1. For these integrations we took a fiducial value for 
the RV of 400 km s$^{-1}$. We used a Monte Carlo method to sample a set of velocities for 
the Magellanic Clouds within the range of their observed errors (listed in Table~\ref{tab:tab2}). 
We took 1,000 samples and YMCA-1 remains bound to the LMC in all cases. There are 
8 integrations (0.8 per cent) in which the cluster eventually escapes the LMC, however this is only 
after at least 3 orbits.

\begin{figure}
\includegraphics[width=\columnwidth]{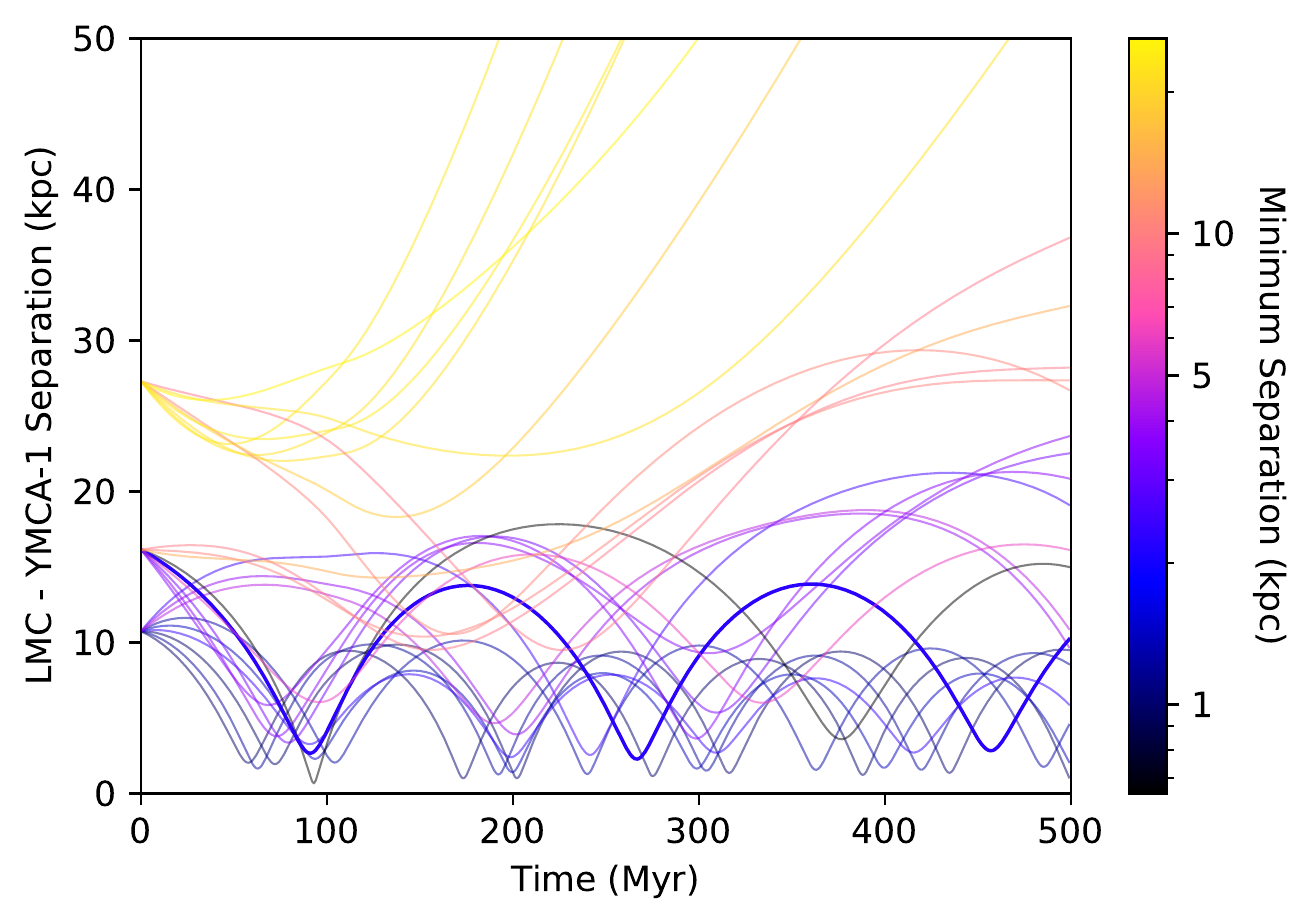}
\caption{Impact of YMCA-1 uncertainties on its orbit. The separation between the LMC 
and YMCA-1 as a function of time for the 27 integrations in which the cluster's distance 
and proper motions are shifted by $-$1, 0, and $+$1 sigma is shown. The lines are coloured
by the minimum approach of the cluster to the LMC within the past 500 Myrs. The fiducial 
orbit is shown in bold (corresponding to distance $=60.9$ kpc, pmra $=1.044$ mas yr$^{-1}$, 
and pmdec $=1.107$ mas yr$^{_1}$).}
\label{fig6}
\end{figure}

We also performed Monte Carlo simulations with respect to the distance and proper motion 
errors of YMCA-1. From the 1,000 samples taken, 587 integrations (58.7 per cent) result in YMCA-1 being bound to the LMC for at least 2 orbits and approaching a distance $<10$ kpc. The 
separation between the LMC and YMCA-1 as a function of time is shown in 
Fig.~\ref{fig6} for a subset of this parameter space. 27 integrations are shown corresponding 
to all combinations of $+$1, 0, and $-$1 sigma in distance, pmra, and pmdec. The lines in 
this figure are coloured by the minimum separation between the LMC and YMCA-1 over the 
past 500 Myr and the fiducial case (0 sigma for all three parameters) is shown as a bold line. 
For larger present-day distances or smaller pmra, the cluster doesn't approach the LMC as 
closely and therefore is less likely to have been bound to the LMC.

\begin{table*}
\caption{Adopted parameter for the MW, LMC, and SMC, respectively.}
\label{tab:tab2}
\begin{tabular}{@{}lccc}\hline
& MW & LMC & SMC \\ \hline
DM Mass (M$_\odot$) & $10^{12}$ & $1.8\times10^{11}$ & $1.9\times10^{10}$ \\
r$_{200}$ (kpc) & 166.1 & 92.72 & 45.22 \\
Hernquist Scale Length (kpc) & 8.6 & 4.6 & 2.2 \\
Disk Mass$^a$ (M$_\odot$) & $5.8\times10^{10}$ & $10^{10}$ & $2\times10^9$ \\
Present-day Position$^b$ (kpc) & $-$ & (-0.76, -41.29, -27.15) & (15.20, -36.68, -42.67) \\
Present-day Velocity$^b$ (km s$^{-1}$) & $-$ & ($-57\pm13$, $-226\pm15$, $221\pm19$)$^c$ & ($18\pm6$, $-179\pm16$, $174\pm13$)$^d$ \\\hline
\end{tabular}

\footnotesize{$^a$ Including both the stellar and gaseous components.

$^b$ In Cartesian coordinates (x,y,z) centered on the MW with the sun located at $(x,y,z)=(-8.1,0,0.02)$ kpc.

$^c$ \citet{kallivayaliletal13}; $^d$ \citet{zivicketal2018}.}
\end{table*}

\section{Conclusions}

We used SMASH and {\it Gaia} EDR3 data to determine the astrophysical
parameters of YMCA-1, a stellar object recently discovered in the
outskirts of the LMC. While \citet{gattoetal2021} suggested that it is
a halo star cluster located at 100 kpc from the Galactic centre, 
\citet{gattoetal2022} confirmed its closer distance, at 55 kpc from
the Sun, although they were inconclusive about its physical nature.
The object caught our attention because of the possibility of being
a witness of the interaction between both Magellanic Clouds.

SMASH data of stars in the YMCA-1 field, once they were properly
treated in order to disentangle member from field stars, revealed
the existence of a small star cluster as assessed from the derived
integrated $M_V$ magnitude and half-light radius using stars with
membership probabilities $>$ 50 per cent. Its CMD was used
to derive its fundamental properties by exploring the parameter space 
using synthetic CMDs through the minimization of  
likelihood functions and obtained parameters uncertainties from
standard bootstrap methods. YMCA-1 turned out to be a
small (435 $\msun$) moderately old (age = 9.6 Gyr), moderately metal-poor 
([Fe/H] = -1.16 dex) star cluster, located at a nearly SMC distance (60.9 kpc) 
from the Sun, at $\sim$ 17.1 kpc to the East from the LMC centre.
{\it Gaia} EDR3 data of four likely red giant members allowed us to
derive the mean cluster proper motions. The brightness and size of
YMCA-1 suggest some resemblance to the recently discovered faint star
clusters in the MW outer halo, although YMCA-1 would not seem to match their age-metallicity 
relationship, nor those of MW globular clusters formed
in-situ or ex-situ, besides the fact that  YMCA-1 is some order of magnitudes
less massive than distant halo globular clusters. We also dismissed
an LMC origin, because the cluster age falls within the well-known
LMC cluster age gap, and its metallicity is marginally consistent with the
LMC age-metallicity relationship. Conversely, the entire set of derived parameters
of YMCA-1 tightly match those of SMC star clusters and field stars.

We performed extensive numerical Monte Carlo simulations in order to
trace its trajectory in the space backward in time. 
By considering the MW, the LMC, and the SMC discs as dynamic point particles embedded 
in radially extended dark matter haloes that experience dynamical friction, we explored 
different RV regimes for YMCA-1 finding that RVs $\gtrsim$ 300 km/s lead to bound orbits 
around the LMC during the past 500 Myrs. Such an orbital configuration
is confirmed when uncertainties in the velocities of the LMC and SMC, as well as 
those in the distance and proper motions of YMCA-1 are included. 
The resulting orbital motions would seem to suggest
that YMCA-1  would not be bound to nor
{would have been born in the MW, and its apparent mismatch with the LMC cluster
population properties would lead to speculate on an SMC origin. Such a
scenario was predicted by \citet{carpinteroetal2013}, making YMCA-1
the first tracked star cluster that could have been stripped by the LMC
during any of its approaching to the SMC.

\section*{Acknowledgements}
We thank the referee for the thorough reading of the manuscript and
timely suggestions to improve it. 

\section{Data availability}

Data used in this work are available upon request to the authors.

\newpage









\bsp	
\label{lastpage}
\end{document}